\documentclass[%
  reprint,
  superscriptaddress,
  amsmath,amssymb,
  aps,
  prd,
  nofootinbib
]{revtex4-2}

\usepackage{graphicx}
\usepackage{dcolumn}
\usepackage{bm}
\usepackage{braket}
\usepackage{booktabs}
\usepackage{array}
\usepackage{float}
\usepackage[colorlinks=true,linkcolor=blue,urlcolor=blue,citecolor=blue]{hyperref}
\usepackage{enumitem}
\usepackage{xcolor}

\begin{document}

\title{Revisiting In-Medium QCD Effects on Spin-Polarized Strange Quark Stars}

\author{Manas Lohani}
\affiliation{Department of Physics, Hansraj College, University of Delhi, Delhi - 110007, India}
\author{Yogesh Kumar}
\email{ykumar@hrc.du.ac.in (Corresponding Author)} 
\affiliation{Department of Physics, Hansraj College, University of Delhi, Delhi - 110007, India}
\author{Poonam Jain}
\email{pjain\_phy@aurobindo.du.ac.in (Corresponding Author)} 
\affiliation{Department of Physics, Sri Aurobindo College, University of Delhi, Delhi - 110017, India}
\author{Om Prakash}
\affiliation{Department of Physics, Sri Aurobindo College, University of Delhi, Delhi - 110017, India}
\author{Sanjay Tyagi}
\affiliation{Department of Physics, Sri Aurobindo College, University of Delhi, Delhi - 110017, India}
\author{Vinod Kumar}
\affiliation{Department of Physics, University of Lucknow, Lucknow, 226007, India}
\author{T. A. Nahool}
\affiliation{Physics Department, Faculty of Science, South Valley University, Qena, Egypt}

\date{\today}

\begin{abstract}
We investigate the properties of exotic Strange Quark Matter (SQM) with spin polarization and the complex configuration of Strange Quark Stars (SQSs) using a phenomenological MIT Bag Model, enhanced by incorporating a QCD-informed running strange quark mass dependent on the chemical potential. The effective mass using quasiparticle approach is utilized to understand the framework of SQM. The resulting Equation of State (EoS) is constructed for two distinct parameter sets, yielding an energy per baryon below the iron limit for stable configurations and thus supporting the Bodmer-Witten-Terazawa hypothesis that SQM could be the actual ground state of exotic matter. This framework is then used to address the Tolman-Oppenheimer-Volkoff (TOV) equations to study the effect of spin polarization on stellar properties. The model predicts that the maximum stellar mass increases with the degree of spin polarization, a result that diverges from previous constant-mass models. Furthermore, the model's predictions for mass, radius, and surface redshift are in excellent agreement with observational constraints for the compact object Vela X-1. This agreement validates our theoretical approach and strengthens the candidacy of Vela X-1 as a Strange Quark Star. Overall, the model results highlight the importance of in-medium QCD effects in describing dense matter.
\end{abstract}

\maketitle

\section{Introduction}
\label{sec:intro}

The study of compact stars offers a unique intersection of nuclear physics, particle physics, and astrophysics, providing a basis to test theories of matter under conditions of extreme density \cite{Baym:2018, Glendenning:2000}. Understanding the composition of these stellar remnants remains one of the biggest issues in modern science. The theoretical framework for this search is Quantum Chromodynamics (QCD), whose phase diagram explores the different states of strongly interacting matter. The heavy-ion experiments probe the high-temperature regime at zero or low-density to study the complex structure of Quark-Gluon Plasma (QGP) \cite{Yogesh:2012, Yogesh:2015, Yogesh:2015a}. On the other part, the high-density and zero or low-temperature region is the exclusive domain of compact stars \cite{Shapiro:1983}.
\\
Recent astronomical observations have offered constraints on highly dense equation of state (EoS). The discovery of massive pulsars, such as PSR J1614-2230 \cite{Arz:2018}, PSR J0348+0432 \cite{Antoniadis:2013}, and the more recent PSR J0740+6620 \cite{Cromartie:2020}, requires that the EoS be stiff enough to support stellar masses of at least two solar masses ($M_\odot$). On the other hand, data from gravitational wave events like GW170817 have placed limits on the tidal deformability of neutron stars, suggesting a relatively soft EoS \cite{Sen:2021}. This issue intensified the search for a comprehensive theoretical understanding of stellar interiors.
\\
A suitable possibility, known as the Bodmer--Witten--Terazawa Strange Matter Hypothesis \cite{Bodmer:1971, Witten:1984, Terazawa:1989}, conjectures that matter composed of up, down, and strange quarks named Strange Quark Matter (SQM), may be the true ground state of matter. If correct, this implies the existence of Strange Quark Stars (SQSs), which would be self-bound by the strong interaction and exhibit macroscopic properties different from their traditional neutron star counterparts \cite{Alcock:1986, Farhi:1984}. This distinguishes SQSs from hybrid stars, which are gravitationally bound neutron stars that possess a quark matter core formed through a hadron-quark phase transition \cite{Sen:2021, Burgio:2002}.
\\
Different forms of phenomenological models such as the MIT bag model \cite{Chodos:1974je}, have been extensively used to describe SQM. However, these models often depend on treating fundamental parameters like the quark masses as constants. A more physically relevant approach must account for in-medium QCD effects, where quark properties are expected to evolve with the energy scale. The concept of a density or chemical potential dependent "running" mass has been explored in various contexts, from early models of confinement \cite{Fowler:1981} to modern quasiparticle \cite{Wen:2013} and density-dependent formalisms \cite{Chu:2014, Peng:1999, Bordbar:2012b}, and is an intrinsic feature of frameworks like the Nambu-Jona-Lasinio (NJL) model \cite{Buballa:2005}.
\\
Building on the spin-polarized SQM framework \cite{Bordbar:2011}, this work introduces a key QCD-informed modification: a running strange quark mass dependent on the in-medium chemical potential. In Section~\ref{sec:theory}, we detail this theoretical framework, analyze the resulting EoS, and investigate the properties of SQSs arising from this model, while focusing on the role of spin polarization. In order to check whether this model can produce astrophysical stars, we work on 
Tolman-Oppenheimer-Volkoff (TOV) equations in Section~\ref{sec:structure} to calculate the stellar mass-radius relations. Finally, in Section~\ref{sec:conclusion}, we explore potential observational signatures by summarizing our findings, highlighting the impact of the running mass, and identifying the compact object in Vela X-1 as a compelling counterpart.

\begin{figure*}[!ht]
    \centering
    \includegraphics[width=8.9cm, height=7cm]{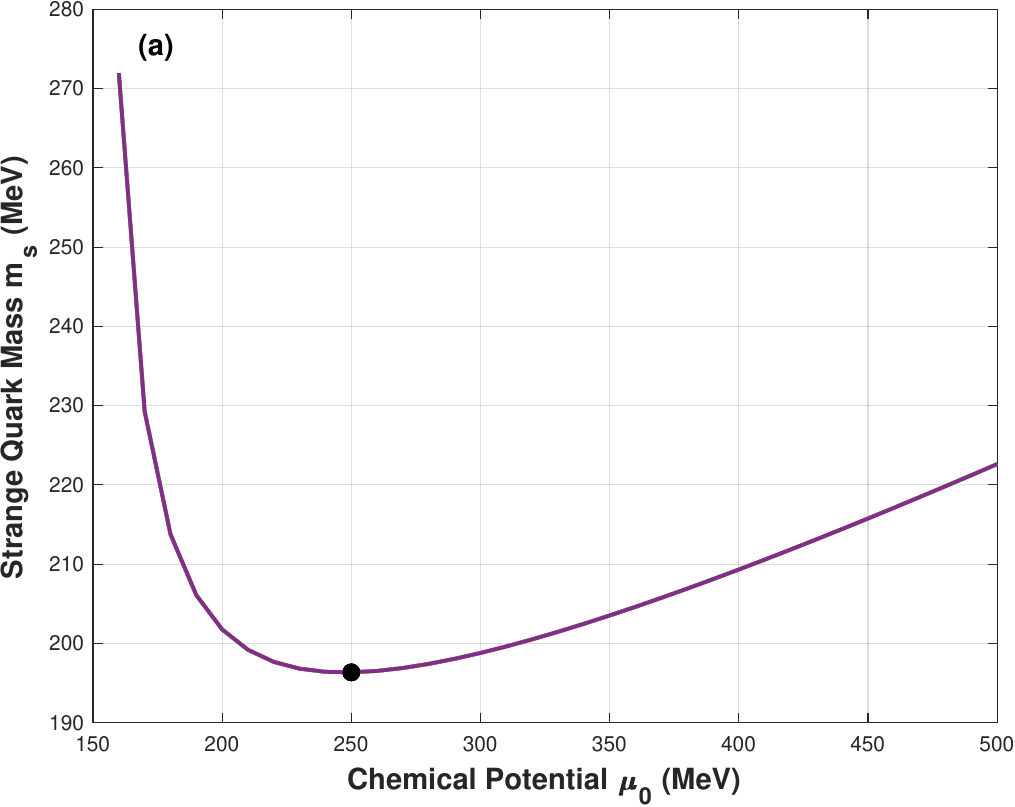}
    \includegraphics[width=8.9cm, height=7cm]{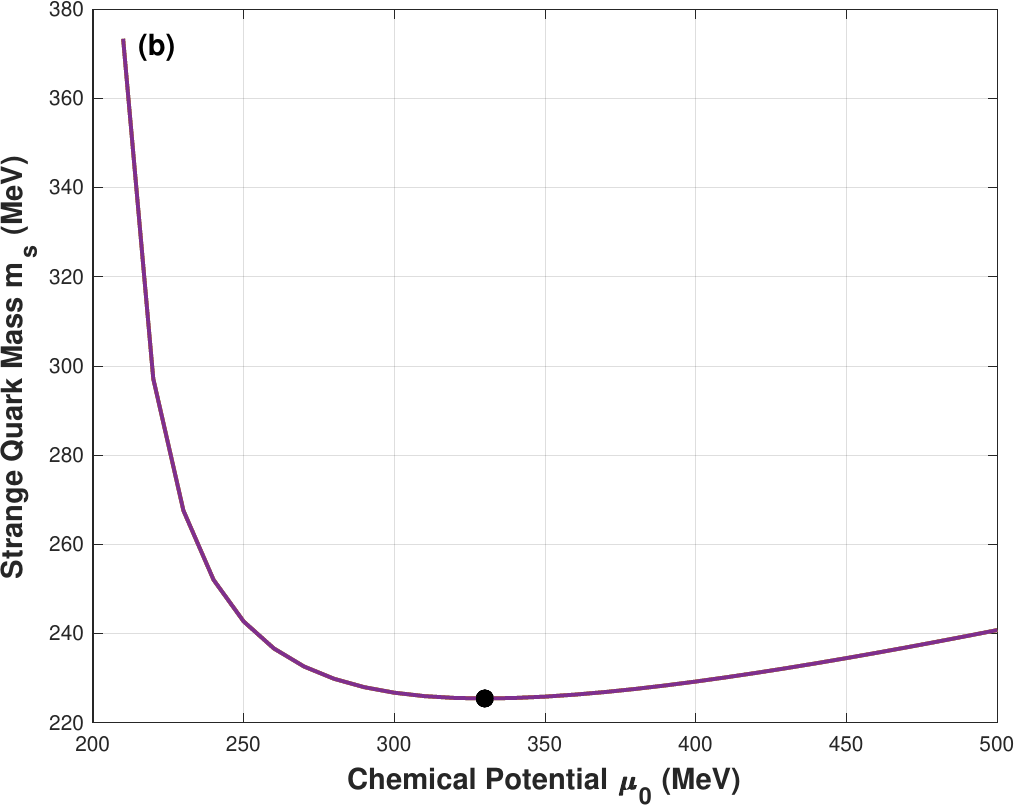}
    \caption{Variation of the running strange quark mass ($m_s$) as a function of chemical potential ($\mu_0$). Panel (a) corresponds to the parameter set ($B_{bag} = 60$ MeV/fm$^3$, $\Lambda_{QCD} = 150$ MeV), while panel (b) shows the result for the set ($B_{bag} = 55$ MeV/fm$^3$, $\Lambda_{QCD} = 200$ MeV). The black dot in each panel indicates the location of the minimum mass.}
    \label{fig:ms_vs_mu}
\end{figure*}

\section{Theoretical Analysis and Brief Model Description of SQM}
\label{sec:theory}
The properties of strange quark matter (SQM) have been investigated by several authors using various methods \cite{Farhi:1984, Witten:1984, Alcock:1986, Alverdyan:2010, Li:2011}. The accelerators at RHIC and LHC provide signals of strangeness enhancement, which has been proposed as an indirect signature of quark-gluon plasma formation and possibly strange quark matter \cite{Weiner:2006}. In this work, we examine spin-polarized SQM consisting of three flavors; $up$ (u), $down$ (d), and $strange$ (s) quarks. These particles are treated as fermions possessing an intrinsic quantum spin that can be oriented "up" ($+$) or "down" ($-$) relative to a quantization axis. The degree of spin alignment is quantified by the polarization parameter \cite{Bordbar:2011}. It is defined as,
\begin{equation*}
    \zeta_i = \frac{\rho_i^{(+)} - \rho_i^{(-)}}{\rho_i},
\end{equation*}
where $\rho_i = \rho_i^{(+)} + \rho_i^{(-)}$ is the total number density for a given quark flavor. While mathematically $\zeta$ can range from -1 (all spins down) to +1 (all spins up), the intrinsic physics is symmetric with respect to the choice of direction. By convention, we take the direction of net polarization as positive value, so the analysis is restricted to the range $0 \leq \zeta \leq 1$ without loss of generality. $\zeta=0$ value is represented as an unpolarized state, while $\zeta=1$ represents a fully spin-aligned state and acts as a ferromagnetic-like state. The primary astrophysical motivation for studying spin polarization comes from the strong magnetic fields in magnetars, which could induce such an alignment \cite{Chu:2014}. Even in the lack of a magnetic field as in our research, analyzing the EoS across a range of $\zeta$ is a crucial theoretical exercise and can provide appreciable results. It allows us to isolate and understand the fundamental impact of spin ordering on the properties of quark matter. However, it provides a baseline for more advanced models. Throughout this work, we assume the simplifying condition of equal polarization for all flavors, i.e., $\zeta_u = \zeta_d = \zeta_s \equiv \zeta$.

To model the SQM, we use the bag value $B_{bag}$ from the MIT Bag Model and the QCD scale parameter, $\Lambda_{QCD}$. In this study, we consider two distinct parameter sets:
\begin{itemize}[leftmargin=*, label={}]
    \item Case I: $B_{bag} = 60$ MeV/fm$^{3}$ and $\Lambda_{QCD} = 150$ MeV,
    \item Case II: $B_{bag} = 55$ MeV/fm$^{3}$ and $\Lambda_{QCD} = 200$ MeV.
\end{itemize}
These parameter sets are chosen to examine the interplay between the confining pressure of the bag and the strength of the quark-quark interaction. A higher value of $\Lambda_{QCD}$ leads to a stronger running coupling constant, providing more self-binding for the quark matter. Consequently, a lower bag constant is sufficient to achieve a stable state with energy per baryon below the iron limit of 930 MeV. This inverse relationship between the required bag constant and the QCD scale parameter allows us to probe different physical scenarios for stable SQM.

\begin{figure*}[!th]
    \centering
    \includegraphics[width=8.9cm, height=7cm]{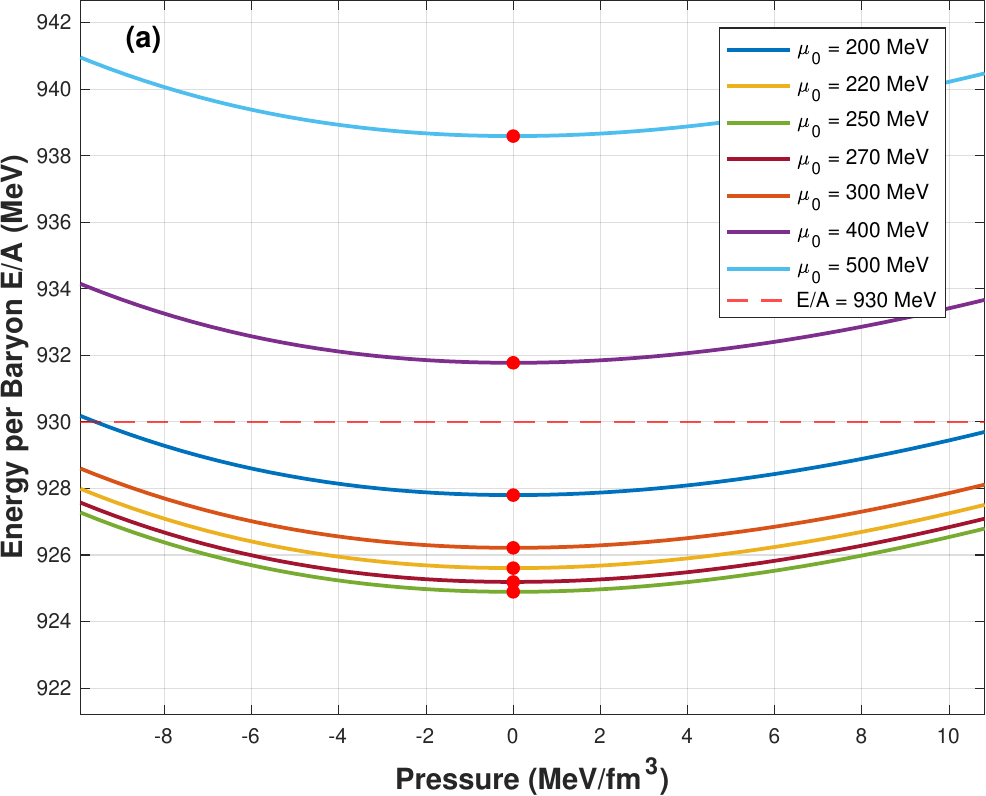}
    \includegraphics[width=8.9cm, height=7cm]{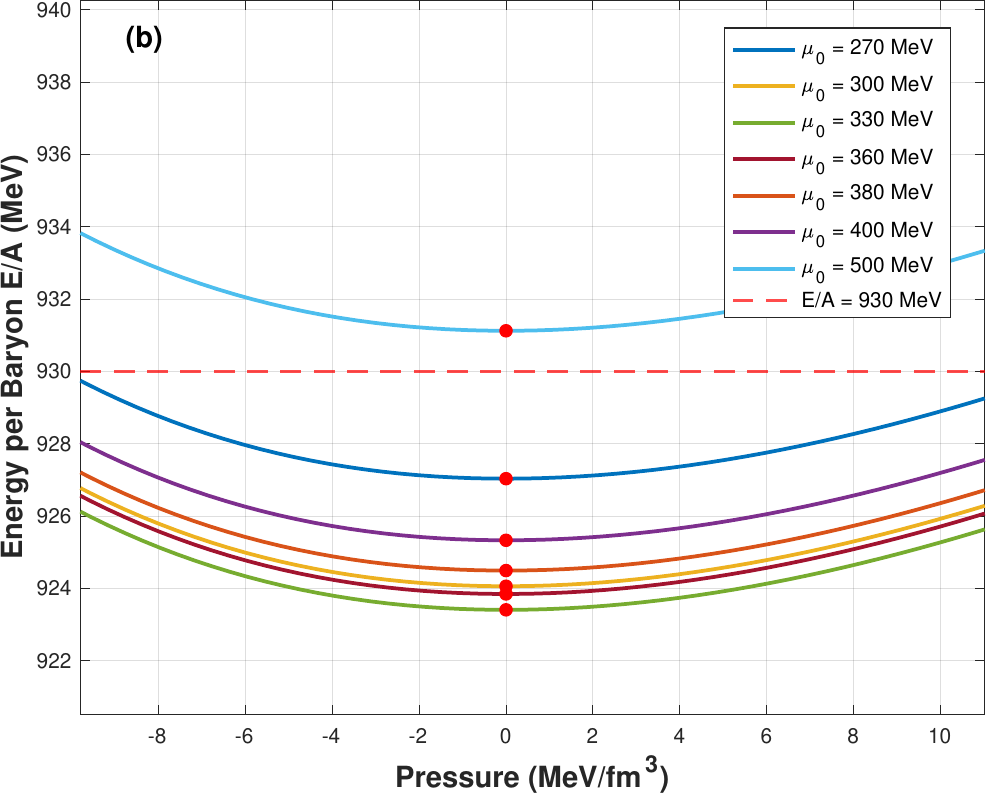}
    \caption{The energy per baryon ($E/A$) as a function of pressure ($P$) for several values of finite chemical potential ($\mu_0$). Panel (a) shows the curves for the parameter set ($B_{bag} = 60$ MeV/fm$^{3}$, $\Lambda_{QCD} = 150$ MeV), and panel (b) corresponds to ($B_{bag} = 55$ MeV/fm$^{3}$, $\Lambda_{QCD} = 200$ MeV). The minimum of each curve occurs at $P=0$. The dashed horizontal line indicates the empirical nuclear saturation point at $E/A=930$ MeV for comparison.}
    \label{fig:EA_vs_P}
\end{figure*}

\begin{table*}[t]
\centering
\large
\caption{Minimum energy per nucleon ($E/A$) as a function of finite chemical potential ($\mu_0$) for two different parameter sets of the bag constant ($B_{bag}$) and QCD scale parameter ($\Lambda_{QCD}$).}
\label{tab:EoS_comparison}
\begin{tabular}{c c | c c}
\toprule
\multicolumn{2}{c|}{$B_{bag} = 60$ MeV/fm$^{3}$ $\Lambda_{QCD} = 150$ MeV} & 
\multicolumn{2}{c}{$B_{bag} = 55$ MeV/fm$^{3}$, $\Lambda_{QCD} = 200$ MeV} \\
\midrule
$\mu_0$ (MeV) & Min. $E/A$ (MeV) & 
$\mu_0$ (MeV) & Min. $E/A$ (MeV) \\
\midrule
200 & 927.8001 & 270 & 927.0368 \\
220 & 925.6056 & 300 & 924.0635 \\
250 & 924.8919 & 330 & 923.4109 \\
270 & 925.1899 & 360 & 923.8498 \\
300 & 926.2144 & 380 & 924.4963 \\
400 & 931.7766 & 400 & 925.3337 \\
500 & 938.5916 & 500 & 931.1257 \\
\bottomrule
\end{tabular}
\end{table*}

Beyond the constant-mass approximation, our model introduces a running mass governed by an independent environmental scale, $\mu_0$. This parameter functions as a phenomenological tool to characterize the dense quark-matter environment. It isolates the qualitative effects of the running mass prescription. We treat $\mu_0$ as an independent input variable rather than a strict thermodynamic chemical potential. This allows us to systematically vary the energy scale and observe the resulting structural changes. This framework is physically motivated by in-medium QCD effects where quark properties are expected to evolve with the energy scale, a feature explored in various forms, including models with dynamical quark masses~\cite{Bordbar:2012}. While we neglect the masses of the up ($u$) and down ($d$) quarks ($m_u=m_d=0$), the strange quark's effective mass squared ($m_s^2$) is adapted from the quasiparticle model formalism and depends on chemical potential~\cite{Srivastava:2010, LinLi:2021, Bannur:2008, Peshier:2000, Kumar:2018}:
\begin{equation}
\label{eq:ms_running}
    m_s^2 = m_{s0}^2 + \sqrt{2} m_{s0} m_q + m_q^2,
\end{equation}
where $m_{s0}$ is the current mass of the strange quark (150 MeV). The term $m_q$ represents the in-medium dynamical mass contribution. Here, we adapt it to be dependent on the chemical potential, representing a leading-order perturbative QCD correction:
\begin{equation*}
    m_q = \sqrt{\frac{g^2 \mu_0^2}{6\pi^2}}.
\end{equation*}
This approach is consistent with other quasiparticle descriptions of dense matter~\cite{Wen:2013, Peshier:2002}. The running coupling constant $g$ is also taken to be dependent on the chemical potential to better reflect QCD dynamics~\cite{Yang:2016}:
\begin{equation}
\label{eq:g_coupling}
    g = \sqrt{\frac{16\pi^2}{(11 - \frac{2n_f}{3}) \log\left(\frac{\mu_0^2}{\Lambda_{QCD}^2}\right)}} 
\end{equation}
In this expression, $n_f$ is the number of active quark flavors. The logarithmic term in the denominator of Equation~(\ref{eq:g_coupling}) enforces the physical condition that $\mu_0$ must be greater than the QCD scale. At chemical potentials just above $\Lambda_{QCD}$, $g$ is large and diverges at $\mu_0 = \Lambda_{QCD}$ where the model is not defined. We define a working range $\Lambda_{QCD} < \mu_0 \leq 500$ MeV for our calculations.

The behaviour of running strange quark mass as a function of chemical potential is shown in Figure~\ref{fig:ms_vs_mu}. The large value of coupling constant $g$ results in a high strange quark mass at chemical potential values close to $\Lambda_{QCD}$. The effect of asymptotic freedom causes $g$ to decrease rapidly as $\mu_0$ increases. This decrease initially dominates over the rising $\mu_0^2$ term in Equation~(\ref{eq:ms_running}), causing the overall mass $m_s$ to fall. This trend continues until a minimum value is reached, which corresponds to the most stable configuration of the system. This minimum, marked by a black dot in each panel, occurs at $\mu_0 = 250$ MeV for Case I and $\mu_0 = 330$ MeV for Case II. After this point, $\mu_0^2$ term increases more rapidly than the slowly changing logarithmic term, so $m_s$ keeps increasing with chemical potential.

Physically, this behaviour reflects the competition between asymptotic freedom and the increasing energy scale of the dense medium. At low chemical potentials, the reduction in the strong interaction strength dominates and lowers the effective strange quark mass. Beyond the minimum, the continued increase in the chemical-potential environment becomes the dominant effect, causing the quasiparticle mass to rise slowly despite the further decrease of the coupling constant.

\begin{figure*}[!th]
    \centering
    \includegraphics[width=16cm, height=5.5cm]{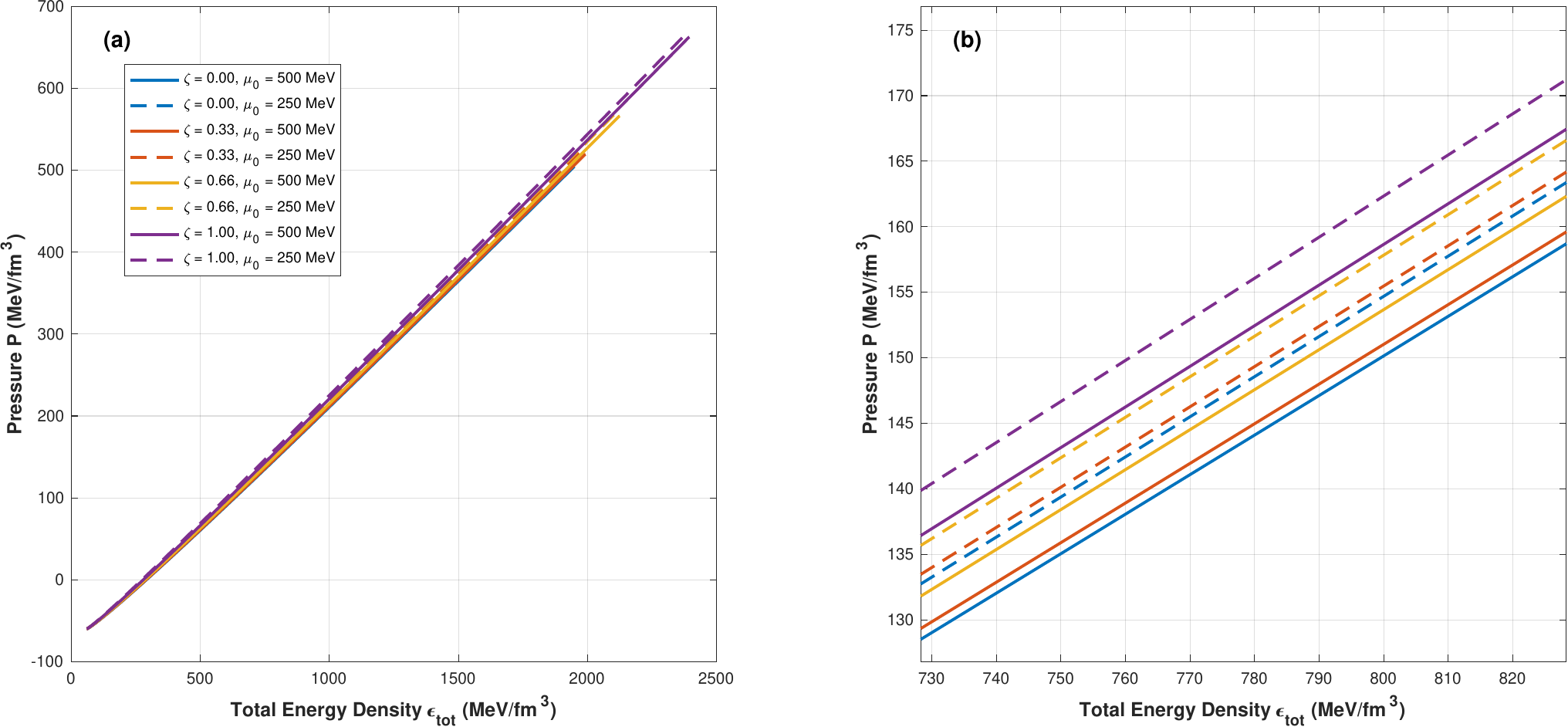}
    \caption{The pressure ($P$) as a function of the total energy density ($\epsilon_{tot}$) for the parameter set ($B_{bag} = 60$ MeV/fm$^{3}$, $\Lambda_{QCD} = 150$ MeV). The plot shows results for different values of the spin polarization ($\zeta$) for the most stable ($\mu_0 = 250$ MeV) and least stable ($\mu_0 = 500$ MeV) configurations. Panel (a) shows the full range, while panel (b) provides a zoomed-in view.}
    \label{fig:P_vs_E_150}
\end{figure*}

\begin{figure*}[!th]
    \centering
    \includegraphics[width=16cm, height=5.5cm]{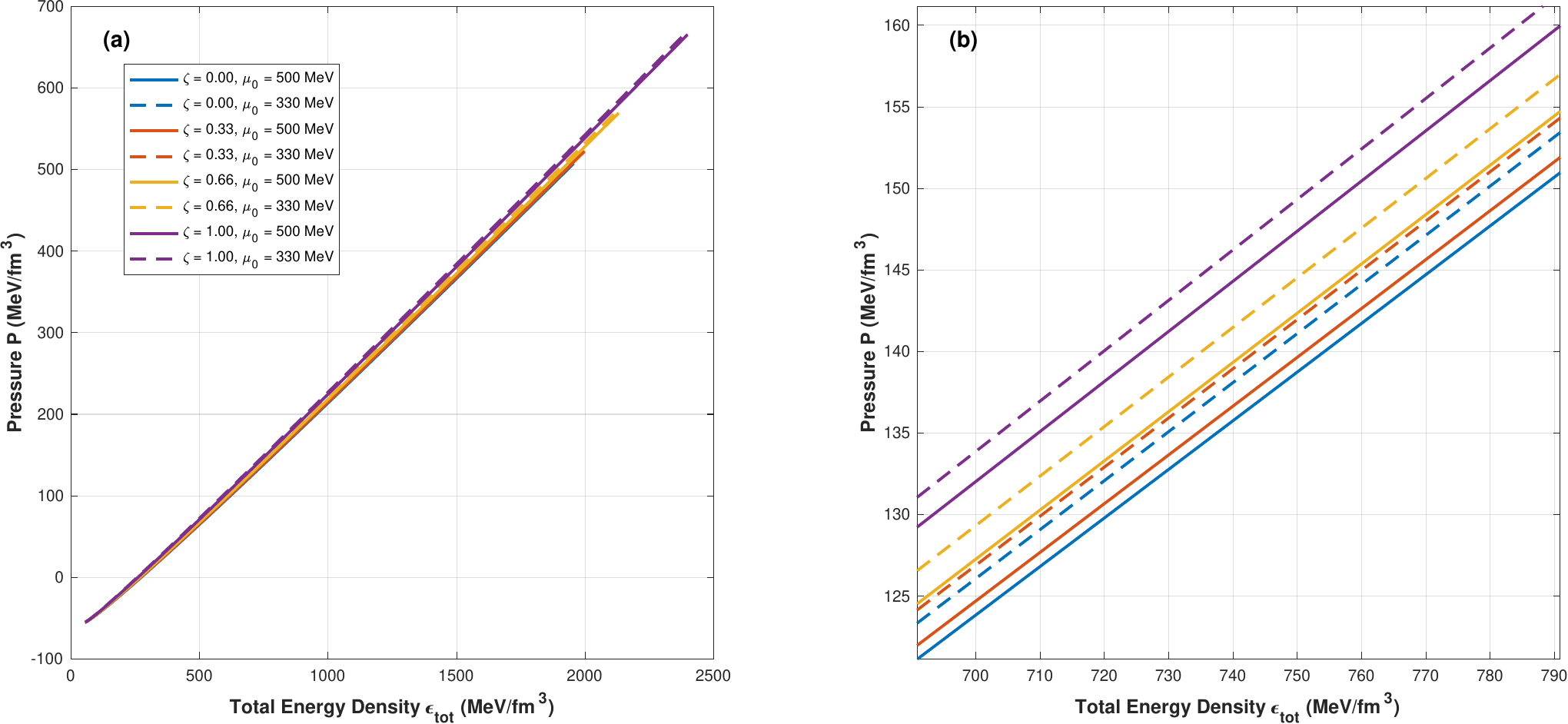}
    \caption{The pressure ($P$) as a function of the total energy density ($\epsilon_{tot}$) for the parameter set ($B_{bag} = 55$ MeV/fm$^{3}$, $\Lambda_{QCD} = 200$ MeV). The plot shows results for different values of the spin polarization ($\zeta$) for the most stable ($\mu_0 = 330$ MeV) and least stable ($\mu_0 = 500$ MeV) configurations. Panel (a) shows the full range, while panel (b) provides a zoomed-in view.}
    \label{fig:P_vs_E_200}
\end{figure*}

\subsection{Equations of State}
\label{sec:eos_calc}
For all subsequent calculations, we adopt the system of natural units where $\hbar = c = 1$, and all physical quantities are expressed in units of MeV. The total energy density ($\epsilon_{tot}$) is calculated from the formalism presented in \cite{Bordbar:2011}. The final analytical expression used in our work is the following.
\begin{equation}
\label{eq:etot_natural}
\begin{split}
    \epsilon_{tot} = & \frac{3}{16\pi^2} \sum_{p=\pm} \left[ k_F^{p} E_F^{p} (2(k_F^{p})^2 + m_s^2) - m_s^4 \ln\left(\frac{k_F^{p} + E_F^{p}}{m_s}\right) \right] \\
    & + \frac{3\pi^{2/3}}{4} \rho^{4/3} \left[ (1+\zeta)^{4/3} + (1-\zeta)^{4/3} \right] + B_{bag},
\end{split}
\end{equation}
where the spin-dependent Fermi momentum, $k_F^{\pm}$, and Fermi energy, $E_F^{\pm}$, are given by:
\begin{equation*}
\label{eq:kF}
    k_F^{\pm} = (\pi^2\rho)^{1/3}(1 \pm \zeta)^{1/3},
\end{equation*}
\begin{equation*}
\label{eq:EF}
    E_F^{\pm} = \sqrt{(k_F^{\pm})^2 + m_s^2}.
\end{equation*}
The pressure ($P$) is then derived from the thermodynamic relation \cite{Glendenning:2000}. It is expressed as:
\begin{equation}
    \label{eq:pressure}
    P(\rho) = \rho \frac{\partial \epsilon_{tot}}{\partial \rho} - \epsilon_{tot}.
\end{equation}
This set of equations very well defines the EoS used in our analysis.

\begin{figure*}[!th]
    \centering
    \includegraphics[width=16cm, height=5.5cm]{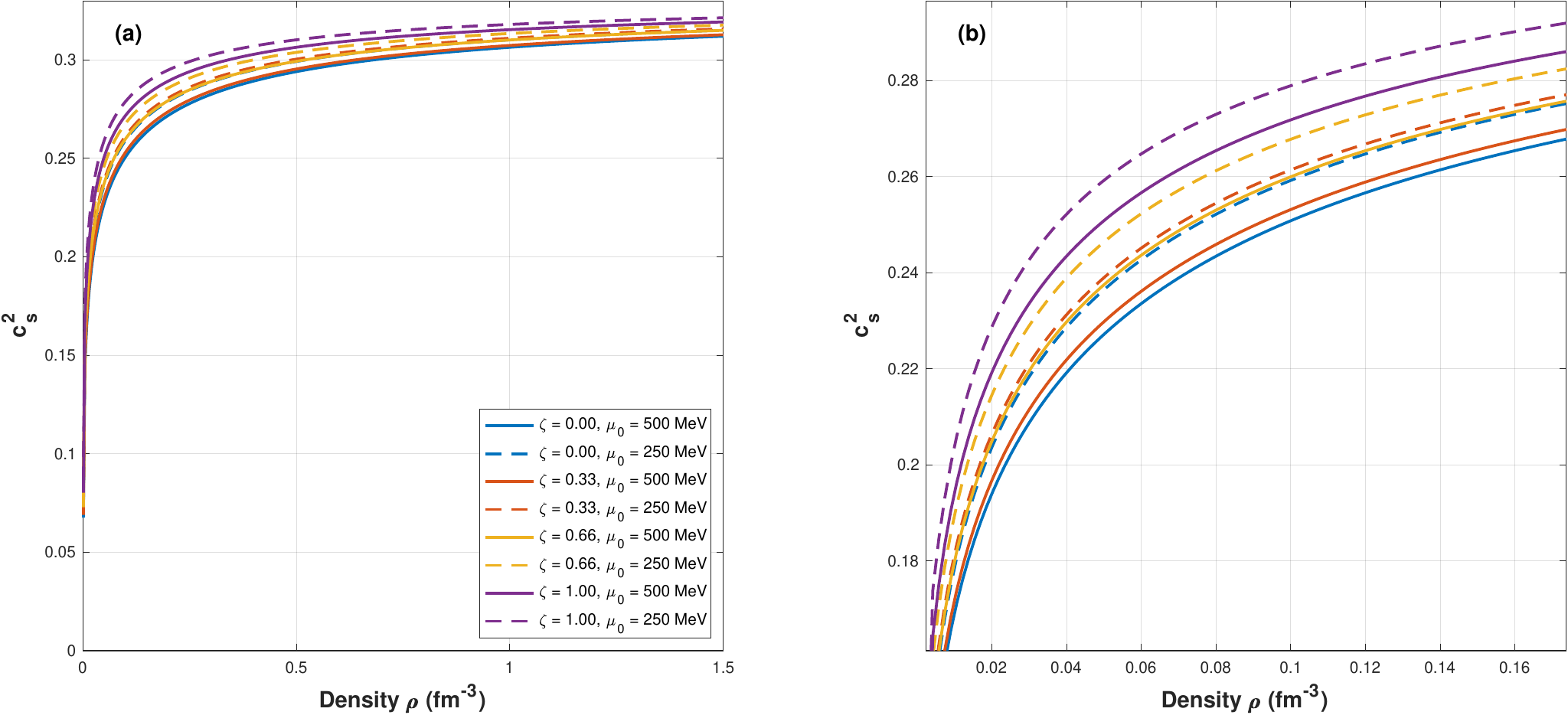}
    \caption{The speed of sound squared ($c_s^2$) as a function of baryonic density ($\rho$) for the parameter set ($B_{bag} = 60$ MeV/fm$^{3}$, $\Lambda_{QCD} = 150$ MeV). The plot shows results for different values of the spin polarization ($\zeta$) for the most stable ($\mu_0 = 250$ MeV) and least stable ($\mu_0 = 500$ MeV) configurations. Panel (a) shows the full density range, while panel (b) provides a zoomed-in view at low densities.}
    \label{fig:cs2_vs_rho_150}
\end{figure*}

\begin{figure*}[!th]
    \centering
    \includegraphics[width=16cm, height=5.5cm]{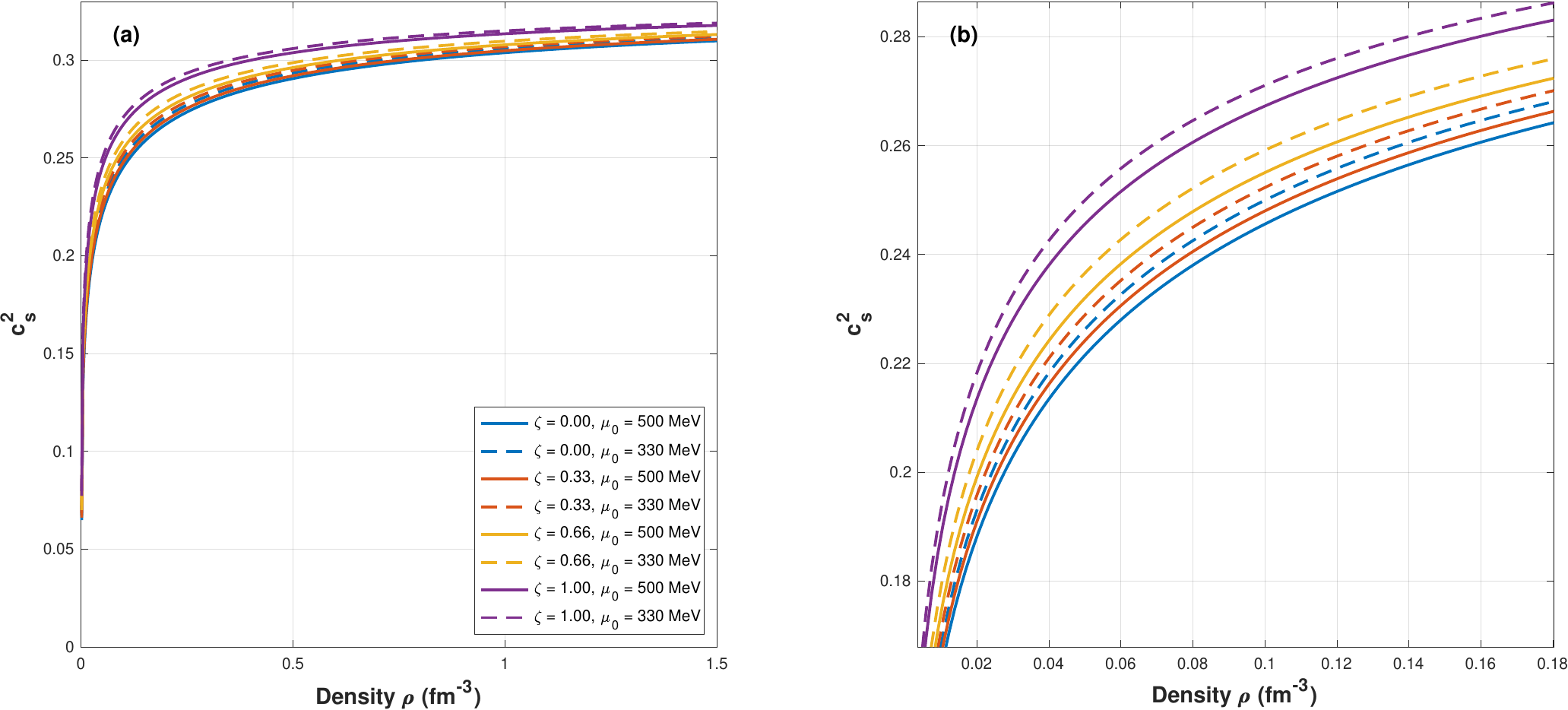}
    \caption{The speed of sound squared ($c_s^2$) as a function of baryonic density ($\rho$) for the parameter set ($B_{bag} = 55$ MeV/fm$^{3}$, $\Lambda_{QCD} = 200$ MeV). The plot shows results for different values of the spin polarization ($\zeta$) for the most stable ($\mu_0 = 330$ MeV) and least stable ($\mu_0 = 500$ MeV) configurations. Panel (a) shows the full density range, while panel (b) provides a zoomed-in view at low densities.}
    \label{fig:cs2_vs_rho_200}
\end{figure*}

\begin{table*}[t!]
\centering
\large
\caption{Maximum value of the speed of sound squared term ($c_s^2$) as a function of the polarization parameter $\zeta$. The values are calculated for two different sets of the bag constant ($B_{bag}$) and QCD scale parameter ($\Lambda_{QCD}$), and are evaluated at their respective stable and unstable chemical potential.}
\label{tab:SoS_zeta_newlayout}
\begin{tabular}{c | c c | c c}
\toprule
& \multicolumn{2}{c|}{$B_{bag} = 60$ MeV/fm$^{3}$, $\Lambda_{QCD} = 150$ MeV} & \multicolumn{2}{c}{$B_{bag} = 55$ MeV/fm$^{3}$, $\Lambda_{QCD} = 200$ MeV} \\
\midrule
& \multicolumn{2}{c|}{Max. $c_s^2$} & \multicolumn{2}{c}{Max. $c_s^2$} \\
$\zeta$ & $\mu_0=250$ MeV & $\mu_0=500$ MeV & $\mu_0=330$ MeV & $\mu_0=500$ MeV \\
\midrule
0.00 & 0.3158 & 0.3128 & 0.3125 & 0.3108 \\
0.33 & 0.3164 & 0.3135 & 0.3132 & 0.3116 \\
0.66 & 0.3182 & 0.3156 & 0.3154 & 0.3139 \\
1.00 & 0.3218 & 0.3198 & 0.3196 & 0.3184 \\
\bottomrule
\end{tabular}
\end{table*}

\subsection{Baryonic Energy and Stability}
\label{sec:stability}
To assess the stability of the system, we calculate the energy per baryon, $E/A$, and the speed of sound squared, $c_s^2$. The energy per baryon is derived from Equation (\ref{eq:pressure}):
\begin{equation}
\label{eq:E_per_A}
    E/A = \frac{\partial \epsilon_{tot}}{\partial \rho} - \frac{P}{\rho}.
\end{equation}
The most stable state of the system is found at the density where $E/A$ reaches its lowest value while the speed of sound squared characterizes the stiffness of the EoS. It is given by the fundamental relation \cite{Glendenning:2000}:
\begin{equation}
\label{eq:cs2}
    c_s^2 = \frac{\partial P}{\partial \epsilon_{tot}}.
\end{equation}
Here, $c_s^2 > 0$ gives mechanical stability, while $c_s^2 \le 1$ is a relativistic constraint that ensures information does not travel faster than light.

In Figure~\ref{fig:EA_vs_P}, we plot the energy per baryon as a function of pressure for various values of chemical potential. We show the specific minimum values for each curve tabulated in Table~\ref{tab:EoS_comparison}. In both Case I and Case II, the minima for most configurations lie far below that of iron considered as the most stable nucleus. Energy per nucleon for Iron is about 930 MeV (dashed line). Results are in support with the work of Bodmer-Witten-Terazawa hypothesis \cite{Bodmer:1971, Witten:1984, Terazawa:1989}. This indicates that the strange quark matter could be regarded as a real ground state of matter and hence poses as an interesting signal for the confirmation of strangelets. We note that the present stability analysis is restricted to three-flavor strange quark matter. A complete determination of the stability window would additionally require an investigation of two-flavor quark matter (ud) to verify its instability with respect to ordinary nuclear matter. We also showed cases in which minimum $E/A$ is above 930 MeV depicting a less stable state. A comparison of the two parameter sets reveals that Case II yields a more deeply bound absolute minimum ($E/A = 923.4109$ MeV) than Case I ($E/A = 924.8919$ MeV). This comparison suggests that the former produces more stable SQM.

Based on the above findings, we select two representative configurations for further analysis: the most stable ($\mu_0 = 250$ and $330$ MeV for Cases I and II, respectively) and the least stable ($\mu_0 = 500$ MeV) configuration. The corresponding equations of state are shown in Figures~\ref{fig:P_vs_E_150} and \ref{fig:P_vs_E_200}. In both parameter sets, the pressure increases monotonically with the total energy density, yielding a physically acceptable equation of state over the entire range considered. For a fixed value of the spin polarization parameter $\zeta$, the stable configurations consistently produce higher pressures at a given energy density than their corresponding unstable counterparts. Furthermore, increasing $\zeta$ systematically shifts the equation of state towards higher pressures. This behaviour indicates that spin polarization enhances the resistance of strange quark matter to compression and therefore leads to a progressively stiffer equation of state. Physically, this trend originates from the increasing imbalance between spin-up and spin-down quark populations. As the degree of polarization increases, the available phase space becomes more restricted, forcing quarks to occupy higher momentum states and thereby increasing the Fermi pressure. The stable configurations exhibit this effect more strongly, which explains their systematically higher pressures compared to the corresponding unstable states.

The stiffening of the equation of state with increasing polarization is astrophysically significant because it enhances the ability of strange quark matter to resist gravitational compression and therefore influences the maximum mass and radius attainable by a strange quark star. The resulting effects on the stellar structure are discussed in Section~\ref{sec:structure}.

The stiffness of the EoS is most precisely characterized by the speed of sound squared ($c_s^2$), shown in Figures~\ref{fig:cs2_vs_rho_150} and \ref{fig:cs2_vs_rho_200}. For both parameter sets, the stable configurations ($\mu_0 = 250$ MeV for Case I and $\mu_0 = 330$ MeV for Case II) consistently exhibit higher values of $c_s^2$ than their corresponding unstable counterparts ($\mu_0 = 500$ MeV) across the entire density range and for all polarization states. This behaviour indicates that the stable configurations possess a stiffer equation of state and are therefore more resistant to compression. The difference becomes increasingly pronounced with increasing spin polarization, reflecting the enhanced contribution of the Fermi pressure in polarized matter. These results are fully consistent with the pressure--energy density relations discussed above, where the stable configurations also produce higher pressures at a given energy density.

For both parameter sets, the plots and the data in Table~\ref{tab:SoS_zeta_newlayout} clearly show that the value of $c_s^2$ increases monotonically with the spin polarization parameter $\zeta$ for any given configuration. This trend is a direct consequence of the Pauli exclusion principle, which enhances the resistance of the matter to compression and consequently increases the speed of sound. The maximum values obtained remain well below the causal limit of $c_s^2 \leq 1$ and approach the conformal limit of $c_s^2 = 1/3$~\cite{Bedaque:2015} as complete polarization ($\zeta \rightarrow 1$) is approached. Therefore, all considered configurations satisfy the fundamental physical constraints while exhibiting the expected stiffening associated with increasing polarization. These results are subsequently employed to model the stellar structure discussed in the next section.

\section{Structure of Strange Quark Star}
\label{sec:structure}

The gravitational mass term ($M$) and radius term ($R$), the key observables of compact stars, are especially significant in astrophysics. Thus, we compute these structural properties for a spin-polarized strange quark star (SQS). Using the EoS derived in the previous section, we can obtain $M$ and $R$ by numerically integrating the general relativistic equations of hydrostatic equilibrium, the Tolman-Oppenheimer-Volkoff (TOV) equations \cite{Shapiro:1983}. In relativistic units, the value of speed of light $c$ is assumed to be 1 and retaining the Gravitational constant $G$ explicitly, the TOV equations are as follows:
\begin{equation}
\label{eq:tov1}
    \frac{dm}{dr} = 4\pi r^2 \epsilon(r),
\end{equation}
\begin{equation}
\label{eq:tov2}
\begin{split}
    \frac{dP}{dr} = & -\frac{G m(r)\epsilon(r)}{r^2} \left( 1 + \frac{P(r)}{\epsilon(r)} \right) \\
    & \times \left( 1 + \frac{4\pi r^3 P(r)}{m(r)} \right) \left( 1 - \frac{2Gm(r)}{r} \right)^{-1},
\end{split}
\end{equation}
where factors $\epsilon(r)$ and $P(r)$ are the energy density value and pressure term at a given radius $r$. The term $m(r)$ is the gravitational mass contained within the radius, as indicated by the integral:
\begin{equation*}
    m(r) = \int_0^r 4\pi r'^2 \epsilon(r') dr'.
\end{equation*}
Here, $r'$ is a dummy integration variable for the radius which is used to distinguish it from the integral's upper limit, $r$.

The integration of the coupled TOV Equations (\ref{eq:tov1}) and ~(\ref{eq:tov2}) begins at the center point of the star where radius is ($r=0$) and proceeds outwards. The process requires a chosen central energy density, $\epsilon_c = \epsilon(r=0)$, as an initial condition. The other boundary conditions are applied at the center with $m(0)=0$ and at the star's surface where the pressure must vanish, i.e., $P(R)=0$. This condition defines the total radius of star, $R$. Next, the star's entire gravitational mass is the mass enclosed at this radius, $M = m(R)$.

Another important astrophysical observable related to the mass factor and radius factor of a compact star is the surface gravitational redshift ($z_s$)~\cite{Glendenning:2000}. It is determined by the star's compactness and is given by the expression:
\begin{equation}
    \label{eq:redshift}
    z_s = \left(1 - \frac{2GM}{R}\right)^{-1/2} - 1.
\end{equation}
We calculate the value of the surface redshift using the fundamental constants and stellar parameters in SI units.

\begin{figure*}[!th]
    \centering
    \includegraphics[width=16cm, height=5.5cm]{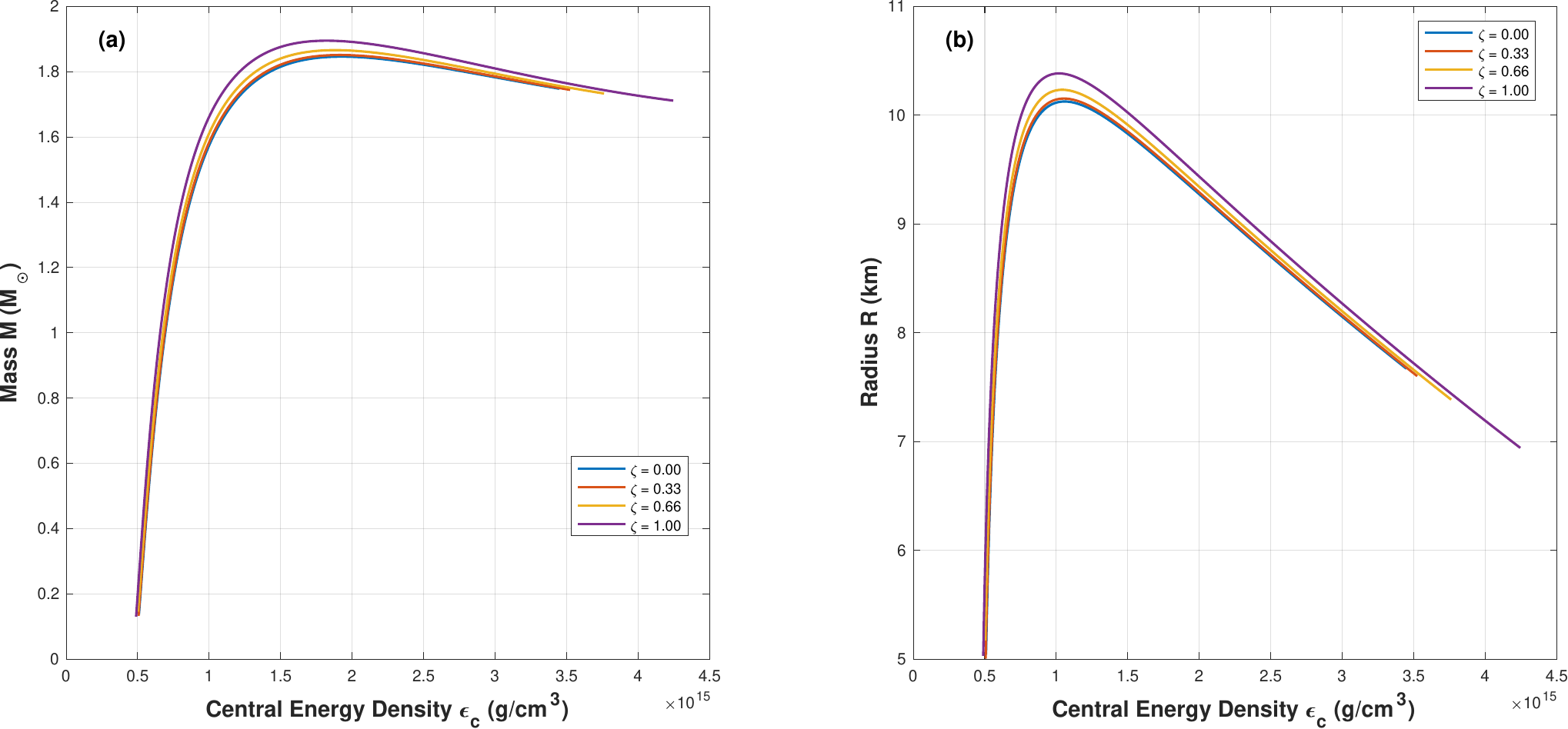}
    \caption{The stellar mass parameter ($M$) in panel (a) and radius factor ($R$) in panel (b) as a function of the central energy density ($\epsilon_c$, in units of $10^{15}\,\mathrm{g\,cm^{-3}}$) for the parameter set ($B_{bag} = 60$ MeV/fm$^{3}$, $\Lambda_{QCD} = 150$ MeV. Each curve corresponds to a distinct value of the spin polarization parameter, $\zeta$.}
    \label{fig:MR_vs_Ec_150}
\end{figure*}

\begin{figure*}[!th]
    \centering
    \includegraphics[width=16cm, height=5.5cm]{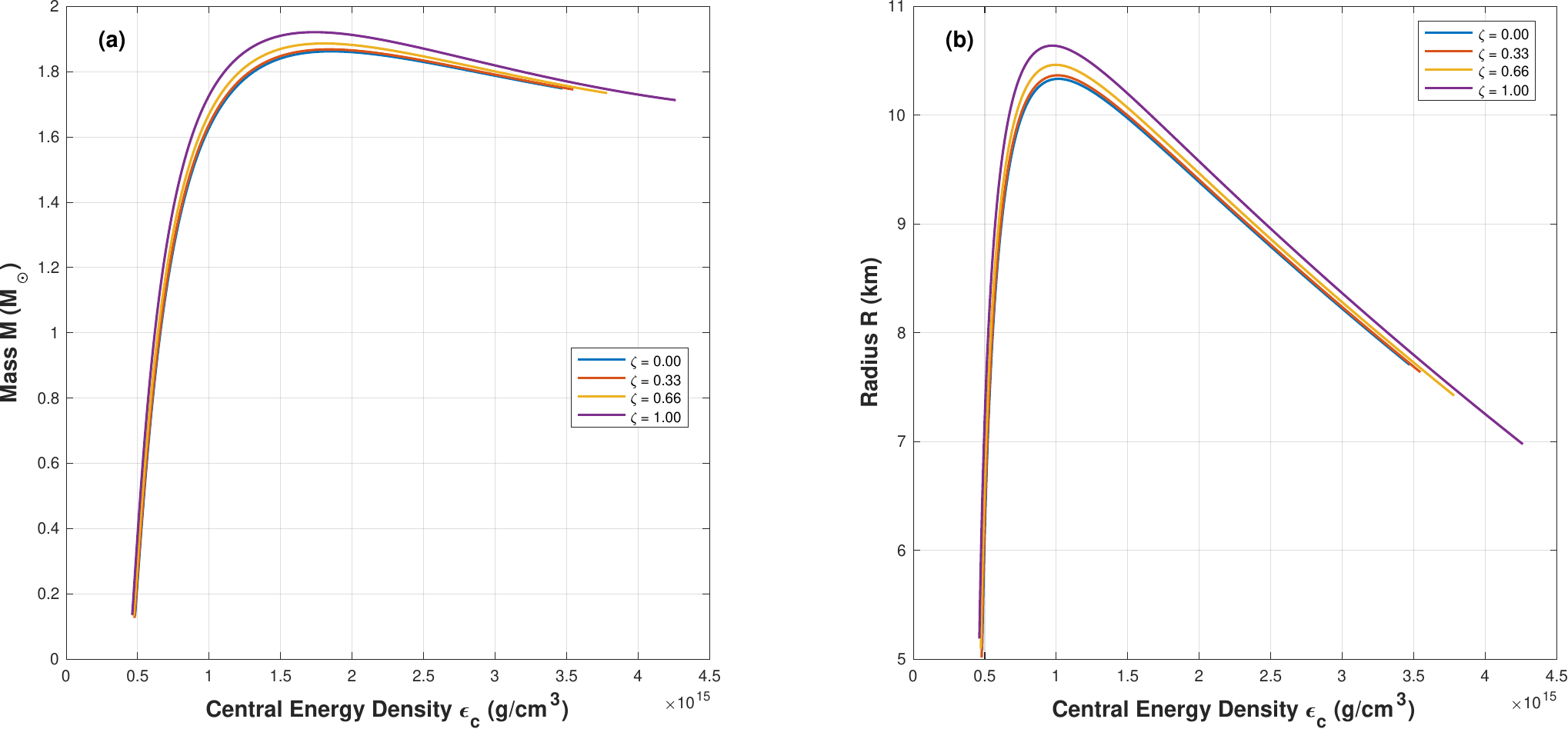}
    \caption{The stellar mass parameter ($M$) in panel (a) and radius factor ($R$) in panel (b) as a function of the central energy density ($\epsilon_c$, in units of $10^{15}\,\mathrm{g\,cm^{-3}}$) for the parameter set ($B_{bag} = 55$ MeV/fm$^{3}$, $\Lambda_{QCD} = 200$ MeV. Each curve corresponds to a distinct value of the spin polarization parameter, $\zeta$.}
    \label{fig:MR_vs_Ec_200}
\end{figure*}

\begin{figure*}[!th]
    \centering
    \includegraphics[width=17.95cm, height=7cm]{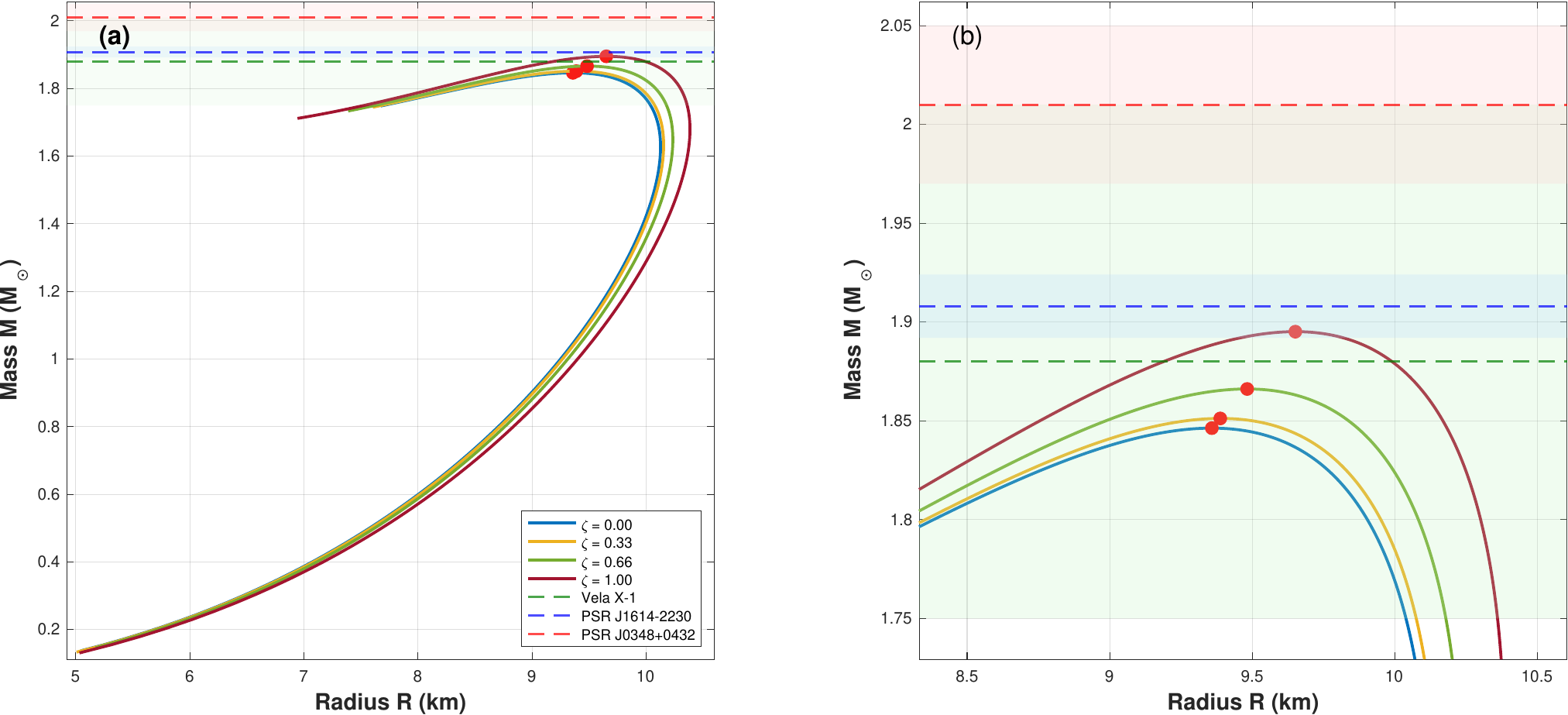}
    \caption{The mass-radius ($M$--$R$) relation for the parameter set ($B_{bag} = 60$ MeV/fm$^{3}$, $\Lambda_{QCD} = 150$ MeV). Each curve corresponds to a distinct value of the spin polarization parameter, $\zeta$. Panel (a) shows the full sequence of stellar configurations, while panel (b) provides a zoomed-in view of the maximum-mass region. The red dots indicate the maximum-mass configuration for each value of $\zeta$. The shaded bands and dashed lines represent observational constraints from selected well-measured compact objects and their associated uncertainties.}
    \label{fig:MR_relation_150}
\end{figure*}

\begin{figure*}[!th]
    \centering
    \includegraphics[width=17.95cm, height=7cm]{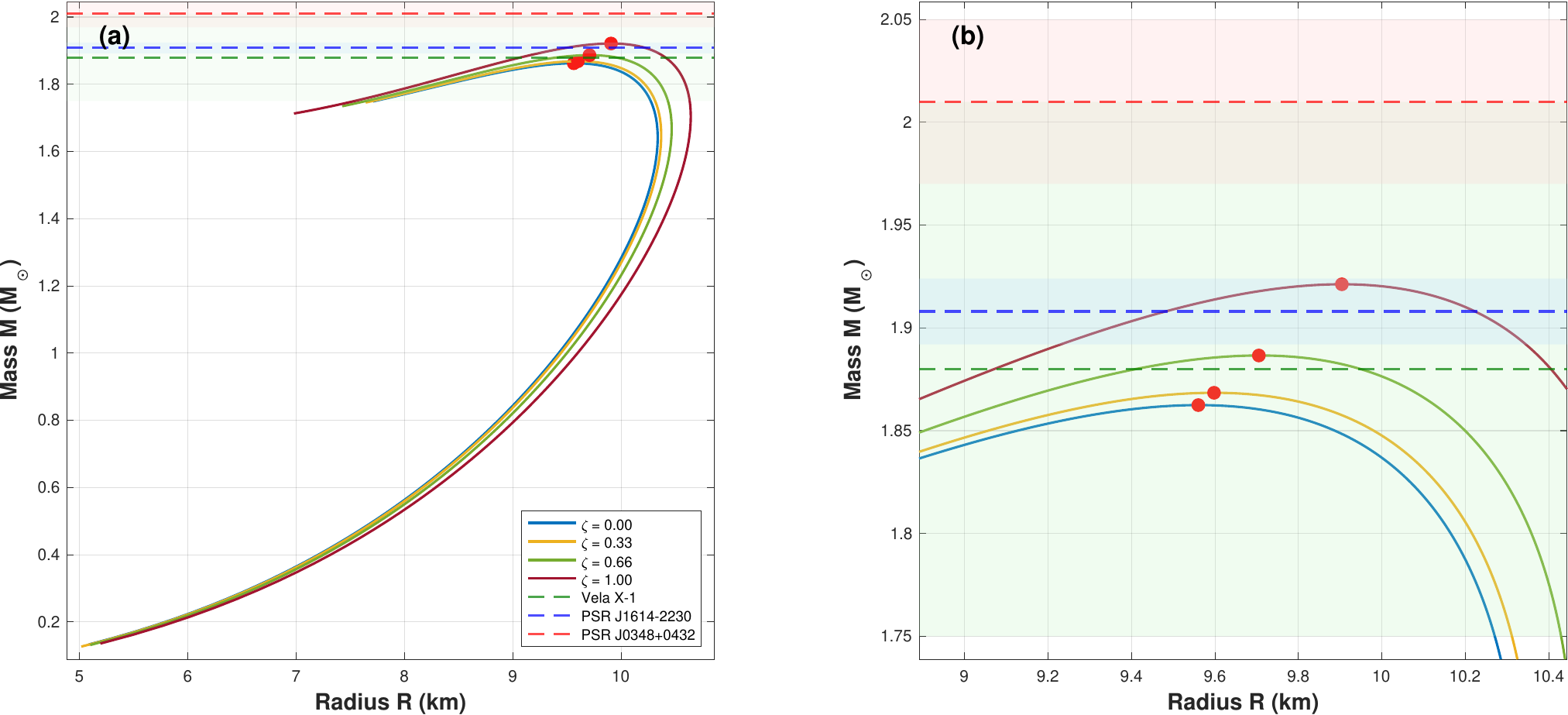}
    \caption{The mass-radius ($M$--$R$) relation for the parameter set ($B_{bag} = 55$ MeV/fm$^{3}$, $\Lambda_{QCD} = 200$ MeV). Each curve corresponds to a distinct value of the spin polarization parameter, $\zeta$. Panel (a) shows the full sequence of stellar configurations, while panel (b) provides a zoomed-in view of the maximum-mass region. The red dots indicate the maximum-mass configuration for each value of $\zeta$. The shaded bands and dashed lines represent observational constraints from selected well-measured compact objects and their associated uncertainties.}
    \label{fig:MR_relation_200}
\end{figure*}

\begin{table*}[t!]
\centering
\large
\caption{Maximum stellar mass ($M_{max}$), the corresponding radius ($R$), and surface redshift ($z_s$) as a function of the polarization parameter $\zeta$. Results are shown for two different sets of the bag constant ($B_{bag}$) and QCD scale parameter ($\Lambda_{QCD}$).}
\label{tab:MRz_relation_zeta}
\begin{tabular}{c | c c c | c c c}
\toprule
& \multicolumn{3}{c|}{$B_{bag} = 60$ MeV/fm$^{3}$, $\Lambda_{QCD} = 150$ MeV} & \multicolumn{3}{c}{$B_{bag} = 55$ MeV/fm$^{3}$, $\Lambda_{QCD}= 200$ MeV} \\
\midrule
$\zeta$ & $M_{max}$ ($M_\odot$)  & R (km) & $z_s$ & $M_{max}$ ($M_\odot$) & R (km) & $z_s$ \\
\midrule
0.00 & 1.8463 & 9.3583 & 0.5479 & 1.8625 & 9.5601 & 0.5346 \\
0.33 & 1.8512 & 9.3879 & 0.5474 & 1.8684 & 9.5978 & 0.5338 \\
0.66 & 1.8661 & 9.4821 & 0.5453 & 1.8866 & 9.7051 & 0.5323 \\
1.00 & 1.8951 & 9.6511 & 0.5429 & 1.9212 & 9.9039 & 0.5302 \\
\bottomrule
\end{tabular}
\end{table*}

In order to investigate the features of stars generated by our EoS, we use only the most stable configuration for each parameter set. These correspond to the chemical potentials $\mu_0 = 250$ MeV for Case I and $\mu_0 = 330$ MeV for Case II. The configurations corresponding to $\mu_0 = 500$ MeV are not considered in the stellar structure calculations. Their minimum energy per baryon exceeds the iron limit and fails the stability condition of the Bodmer-Witten-Terazawa hypothesis. Such unstable configurations do not produce viable strange quark stars. These specific cases are included in the EoS analysis solely for comparison with the stable states. We repeat the numerical integration for 1000 different values of the central energy density $\epsilon_c$ for each EoS. This process generates a corresponding family of unique stellar configurations. The resulting stellar mass in units of solar mass $M_{\odot}$ and the radius $R$ are displayed as a function of $\epsilon_c$ in Figures~\ref{fig:MR_vs_Ec_150} and \ref{fig:MR_vs_Ec_200}. These plots demonstrate that both stellar mass and radius initially increase with central density. A maximum mass is eventually reached. Further increases in $\epsilon_c$ cause the configurations to become unstable to gravitational collapse. This instability causes the stellar radius to decrease~\cite{Shapiro:1983}.

The final mass-radius ($M-R$) relations are displayed in Figures~\ref{fig:MR_relation_150} and \ref{fig:MR_relation_200}. From the speed of sound analysis, we noticed that increasing the spin polarization parameter $\zeta$ makes the equation of state stiffer. A stiffer EoS provides more pressure to oppose gravity, thus allowing the star to support a larger maximum mass. This effect is clearly visible in the figures, where the maximum mass point on each curve (indicated by red dots) increases with $\zeta$. The corresponding radius at the maximum mass also increases with polarization. This result is particularly noteworthy as it contrasts with the findings of Bordbar and Peivand work \cite{Bordbar:2011} where increasing polarization was found to decrease the maximum stellar mass. 

This divergence can be understood from the different treatment of the equation of state in the two approaches. In the work of Bordbar and Peivand \cite{Bordbar:2011}, the stellar structure calculations were performed using a constant-mass equation of state, where the increase in energy density associated with polarization appears to outweigh the additional pressure support, leading to lower maximum masses. In the present work, the running strange quark mass allows the identification of the most stable strange quark matter configuration through the energy per baryon analysis before constructing the stellar sequences. For these stable configurations, the stiffening of the equation of state with increasing polarization becomes the dominant effect. Consequently, the pressure support generated by polarization exceeds the corresponding increase in self-gravity, resulting in larger maximum masses and radii. This behaviour highlights the importance of incorporating both stability considerations and in-medium QCD effects when studying strange quark stars.

To facilitate comparison with observations, representative constraints from several well-characterized compact objects are included in Figures~\ref{fig:MR_relation_150} and \ref{fig:MR_relation_200}. The green shaded region centered on the green dashed line corresponds to Vela X-1, for which a representative mass of $M=1.88\pm0.13 M_{\odot}$ was adopted~\cite{Quaintrell:2003}. The blue shaded region corresponds to PSR J1614$-$2230 with a measured mass of $M=1.908\pm0.016 M_{\odot}$~\cite{Arz:2018}, while the red shaded region represents PSR J0348$+$0432 with a measured mass of $M=2.01\pm0.04 M_{\odot}$~\cite{Antoniadis:2013}. In each case, the horizontal dashed line denotes the central observational value, whereas the surrounding shaded band represents the associated observational uncertainty. These constraints allow a direct visual comparison between the predicted maximum masses of the theoretical stellar sequences and the masses inferred from precision measurements of compact stars.

The calculated stellar configurations are consistent with the observational range inferred for Vela X-1 and remain compatible with the uncertainty interval associated with PSR J1614$-$2230, particularly for larger values of the polarization parameter. In contrast, the present model is unable to reproduce the observed mass of PSR J0348$+$0432.

Additional observational constraints arise from NICER measurements of PSR J0030$+$0451 and PSR J0740$+$6620. For PSR J0030$+$0451, NICER analyses indicate a mass of approximately $1.44 M_{\odot}$ and a radius close to $13$ km~\cite{Miller:2019}. Similarly, PSR J0740$+$6620 has a measured mass exceeding $2 M_{\odot}$ together with a radius of approximately $12$--$14$ km~\cite{Cromartie:2020,Miller:2021,Riley:2021}. These objects are not displayed in the mass-radius diagrams because their inferred radii lie outside the radius range predicted by the present model. The compact object associated with GW190814 provides an even stronger constraint, with an inferred mass in the range $2.50$--$2.67 M_{\odot}$, which is substantially larger than the maximum masses obtained in the present work.

The relatively small radii predicted by the present model are characteristic of compact self-bound strange quark stars and remain below those typically obtained from hadronic and hybrid-star equations of state~\cite{Sen:2021,Sen:2018}. Among the observational candidates considered, Vela X-1 shows the closest correspondence with the present calculations. Multiple independent studies have reported a mass in the range $1.8$--$1.9 M_{\odot}$ for this system~\cite{Barziv:2001,Quaintrell:2003,Rawls:2011}. Adopting the representative values $M=1.88 M_{\odot}$ and $R=9.56$ km gives a surface gravitational redshift of approximately $z_s \approx 0.544$, which lies within the range predicted in Table~\ref{tab:MRz_relation_zeta}. The agreement in mass, radius, and redshift supports Vela X-1 as a possible strange quark star candidate within the framework of the present model.

\section{Summary and Conclusion}
\label{sec:conclusion}

In this study, we have examined the characteristics of strange quark matter with spin polarization and the corresponding structure of strange quark stars. In Section~\ref{sec:theory}, we detailed our theoretical framework, starting from a modified MIT Bag Model where we introduced a running strange quark mass dependent on chemical potential via Equation~(\ref{eq:ms_running}) and Equation~(\ref{eq:g_coupling}). Then we defined the equation of state in Subsection~\ref{sec:eos_calc} by calculating the energy density and pressure for various spin polarizations using Equation~(\ref{eq:etot_natural}) and Equation~(\ref{eq:pressure}). By analyzing the energy per baryon, calculated with Equation~(\ref{eq:E_per_A}), we identified the most stable and least stable configurations for each case, as summarized in Table~\ref{tab:EoS_comparison}. The EoS characteristics and the speed of sound calculated from Equation~(\ref{eq:cs2}) and shown in Figures~(\ref{fig:cs2_vs_rho_150}-\ref{fig:cs2_vs_rho_200}) help us in confirming that the model respects the relativistic constraint and providing insight into the stiffness of the EoS (Table~\ref{tab:SoS_zeta_newlayout}). Subsequently, in Section~\ref{sec:structure}, the Tolman-Oppenheimer-Volkoff (TOV) equations were solved to determine the macroscopic properties of the resulting stars, presenting the mass-radius relations in Figures~\ref{fig:MR_relation_150} and \ref{fig:MR_relation_200}. Our numerical approach proved to be efficient and successfully achieved the goal of determining a valid range for the maximum mass, radius, and surface redshift (calculated with Equation~(\ref{eq:redshift})) of SQSs. The results are summarized in Table~\ref{tab:MRz_relation_zeta}.

Our analysis demonstrates a distinct novelty for theoreticians utilising the quasiparticle model. The introduction of a chemical-potential-dependent running strange quark mass alters the stability and macroscopic properties of strange quark matter. This phenomenological model isolates the effects of a running mass alongside spin polarisation. The resulting stiffening of the equation of state with increasing polarisation is astrophysically relevant. It improves the ability of strange quark matter to resist gravitational compression. This influences the maximum mass and radius attainable by a strange quark star. This behaviour highlights the necessity of incorporating stability considerations and in-medium QCD effects. The agreement in mass, radius, and redshift supports Vela X-1 as a strange quark star candidate. This validates the theoretical framework and provides a foundation to construct more comprehensive astrophysical models.

The energy-per-baryon analysis identified distinct stability windows for the two parameter sets considered. The most stable configurations occurred at $\mu_0=250$ MeV and $\mu_0=330$ MeV for Case I and Case II respectively. The minimum energy per baryon was found to lie below the iron benchmark of $930$ MeV. This supports the Bodmer-Witten-Terazawa hypothesis for the existence of absolutely stable strange quark matter. The speed of sound analysis revealed that increasing the spin polarisation parameter stiffens the equation of state while respecting the relativistic causality constraint. In contrast to previous studies employing a constant strange-quark mass, the present model predicts an increase in the maximum mass with polarisation. The combined effects of the running-mass prescription and the stability analysis modify the balance between pressure support and self-gravity. This allows the polarisation-induced stiffening to become the dominant effect.

Comparison with observational constraints indicates that the predicted stellar configurations remain consistent with the measured mass range of PSR J1614-2230. More massive objects such as PSR J0348+0432, PSR J0740+6620, and the secondary component of GW190814 are not reproduced within the present parameter space. This indicates that the resulting equation of state remains relatively soft despite the stiffening induced by polarisation. These results suggest that the model favours compact self-bound strange quark stars with relatively small radii and moderate maximum masses.

The present framework serves as a strong foundational step that deliberately isolates the thermodynamic effects of a running mass and spin polarisation. Future extensions can build upon this baseline to investigate matter under broader astrophysical conditions. For instance, integrating strong magnetic fields would enable the study of Landau quantization and magnetic-field-induced spin alignment. Subsequent models can also incorporate beta-equilibrium and charge neutrality conditions to examine the resulting flavour asymmetry and its impact on the particle composition. Furthermore, the present calculations employ a standard phenomenological strange quark mass of $150$ MeV, which is conventionally used in MIT Bag Model studies. A systematic investigation utilising alternative mass constraints could offer additional insight into the stability window. Finally, advanced quasiparticle approaches or the NJL model can be explored to gain a deeper understanding of the phase structure at high density.

\section*{Acknowledgements}

The Department of Physics at Hansraj College, University of Delhi, and other institutes provided the necessary research facilities and a favorable atmosphere needed to complete high energy physics work, for which the authors are always thankful.

\end{document}